\RequirePackage{lineno}
\documentclass[twocolumn,showpacs,aps,prl,superscriptaddress,floatfix]{revtex4-1}
\usepackage{graphicx}  
\usepackage{dcolumn}   
\usepackage{bm}        
\usepackage{amssymb}   
\usepackage{amsfonts}  
\usepackage{hyperref}
\usepackage{amsmath}
\usepackage{mathtools}
\usepackage{xspace}
\usepackage{gensymb}
\usepackage{mathrsfs}
\def \DzDzbar  {D^{0}\bar{D^{0}}}
\def \DDbar    {D\bar{D}}

\def \Dz       {D^{0}}

\def \Kp       {K^+}
\def \Km       {K^-}

\def \pip      {\pi^+}
\def \pim      {\pi^-}
\def \piz      {\pi^0}

\def \psipp    {\psi(3770)}

\def \De       {\Delta E}
\def \Ebeam    {E_{\rm beam}}
\def \pppm     {\pi^+\pi^-}

\def \gev  {\mbox{GeV}}

\def \gevcc{\mbox{GeV/$c^2$}}

\def \mevcc{\mbox{MeV/$c^2$}}
\def \ifb  {\mbox{fb$^{-1}$}}

\def \dz      {D^{0}}

\def \mbc     {M_{\rm BC}}

\def \ks    {K_{S}^{0}}

\def \gev  {\mbox{GeV}}

\def \gevcc{\mbox{GeV/$c^2$}}

\def \mevcc{\mbox{MeV/$c^2$}}
\def \ifb  {\mbox{fb$^{-1}$}}

\def \epem {e^+e^-}

\def \pppm {\pi^{+}\pi^{-}}
\def \piz  {\pi^0}
\def \pip  {\pi^+}
\def \pim  {\pi^-}

\def\simge{\mathrel{
   \rlap{\raise 0.511ex \hbox{$>$}}{\lower 0.511ex \hbox{$\sim$}}}}
\def\simle{\mathrel{
   \rlap{\raise 0.511ex \hbox{$<$}}{\lower 0.511ex \hbox{$\sim$}}}}

\let\oldequation\equation
\let\oldendequation\endequation

\renewenvironment{equation}
  {\linenomathNonumbers\oldequation}
  {\oldendequation\endlinenomath}

\begin{document}
\title{\bf First measurement of polarizations in the decay \boldmath{$D^0 \to \omega \phi$}}

\author{
		\begin{small}
		\begin{center}
M.~Ablikim$^{1}$, M.~N.~Achasov$^{10,c}$, P.~Adlarson$^{67}$, S. ~Ahmed$^{15}$, M.~Albrecht$^{4}$, R.~Aliberti$^{28}$, A.~Amoroso$^{66A,66C}$, M.~R.~An$^{32}$, Q.~An$^{63,49}$, X.~H.~Bai$^{57}$
% ...
	\\
		\vspace{0.2cm}
		(BESIII Collaboration)\\
		\vspace{0.2cm} {\it
$^{1}$ Institute of High Energy Physics, Beijing 100049, People's Republic of China\\
% ...
$^{4}$ Bochum Ruhr-University, D-44780 Bochum, Germany\\
% ...
$^{10}$ G.I. Budker Institute of Nuclear Physics SB RAS (BINP), Novosibirsk 630090, Russia\\
% ...
$^{15}$ Helmholtz Institute Mainz, Staudinger Weg 18, D-55099 Mainz, Germany\\
% ...
$^{28}$ Johannes Gutenberg University of Mainz, Johann-Joachim-Becher-Weg 45, D-55099 Mainz, Germany\\
% ...
$^{32}$ Liaoning Normal University, Dalian 116029, People's Republic of China\\
% ...
$^{49}$ State Key Laboratory of Particle Detection and Electronics, Beijing 100049, Hefei 230026, People's Republic of China\\
% ...
$^{57}$ University of Jinan, Jinan 250022, People's Republic of China\\
% ...
$^{63}$ University of Science and Technology of China, Hefei 230026, People's Republic of China\\
% ...
$^{66}$ University of Turin and INFN, (A)University of Turin, I-10125, Turin, Italy; (B)University of Eastern Piedmont, I-15121, Alessandria, Italy; (C)INFN, I-10125, Turin, Italy\\
$^{67}$ Uppsala University, Box 516, SE-75120 Uppsala, Sweden\\
% ...
$^{c}$ Also at the Novosibirsk State University, Novosibirsk, 630090, Russia\\
% ...
}\end{center}
		\vspace{0.4cm}
		\end{small}
}

\begin{abstract}
Using a data sample corresponding to an integrated luminosity of 2.93~\ifb\ collected at a center-of-mass energy $\sqrt{s}=3.773~\gev$ by the BESIII detector, the decay $\dz\to\omega\phi$ is observed for the first time. The branching fraction is measured to be $(6.48 \pm 0.96 \pm 0.40)\times 10^{-4}$ with a significance of $6.3 \sigma$, where the first and second uncertainties are statistical and systematic, respectively. An angular analysis reveals that the $\phi$ and $\omega$ mesons from the $\dz \to \omega \phi$ decay are transversely polarized. The  $95\%$ confidence level upper limit on longitudinal polarization fraction is set to be less than $0.24$, which 
is inconsistent with current theoretical expectations and challenges our understanding of the underlying dynamics in charm meson decays.
\end{abstract}

\maketitle

Understanding the long-distance contributions to $D^0-\bar{D}^0$ mixing is crucial in tests of the Standard Model (SM)~\cite{Cheng}.  These contributions arise in two-body hadronic decays of the $D^0$ meson, such as when the $D^0$ meson decays to two vector ($V$) mesons in the process $D^0 \to VV$, which is expected to account for $10\%$ of the total $D^0$ decay width~\cite{Cheng}. 
In constrast to scalar and pseudoscalar mesons, vector mesons can be produced in three polarization states. Therefore, the decay $\dz\to VV$ produces a longitudinal partial-wave amplitude~($H_{0}$), which is $CP$-even, and two transverse partial-wave amplitudes~($H_{\pm}$), which are superpositions of $CP$-even and $CP$-odd states.
The polarization in the $D^0 \to VV$ decay is sensitive to the $V$-$A$ structure of electroweak interactions in the SM, spin correlations, and  final state interactions, among other effects~\cite{Isard, Valencia, Gritsan}.  Thus, in addition to the partial decay widths, the polarization is an interesting observable.

In the last two decades, a ``polarization puzzle'' has arisen in the decays of heavy mesons to two vectors.
In the beauty sector, naive power counting predicts that $B \to VV$ decays are dominated by longitudinal polarization, since the transverse polarization amplitude suffers from a helicity flip suppression on the order of $\Lambda_{\rm QCD}/m_b$. 
The $B$ factories confirmed this prediction in the decay modes $B^0\to\rho^+\rho^-$, $B^+\to \rho^0\rho^+$, and $B^+ \to \rho^0 K^{*+}$, each of which were found to favor the longitudinal configuration~\cite{Baubert1,Baubert2,JZhang}. 
However, in contrast,
the longitudinal amplitude does not dominate in the decay $B \to \phi K^*(892)$, where it provides only 50\% of the rate~\cite{Baubert1,KFChen}.
The observed deviation in this case can be explained either by penguin annihilation~\cite{Kagan,ZTZou}, rescattering of the final-state interactions~\cite{Colangelo,MLadisa}, or new physics beyond the SM model~\cite{EAlvarez,PKDas,CHChen,YDYang,CSHuang}. 
In the charm sector, the situation is more complicated since the heavy quark expansion method is less reliable in this quark mass regime. 
The naive factorization model~\cite{polarization} and the Lorentz invariant-based symmetry model~\cite{Gudrun} predict the longitudinal polarization fraction, $f_L=H_0^2/(H_0^2+H_-^2+H_+^2)$, to be $\sim$0.5 and $0.33$ for $\Dz\to VV$ , respectively. 
However, a previous measurement 
of the decay $\Dz\to \bar{K}^{*0}\rho^0$ actually shows transverse polarization dominates~\cite{mark}, though the measurement suffers from a large uncertainty.   In addition, the most precise measurement of the decay $D^0\to \rho^0\rho^0$ by the FOCUS collaboration shows large longitudinal polarization with $f_L=(71\pm4\pm2)\%$~\cite{focus}.

To date, all experimential measurements of the helicity in  $\Dz\to VV$ decays have been contrary to theoretical predictions, and the puzzle needs to be confirmed with more precise measurements, and validated using more decay modes.
The singly-Cabibbo-suppressed decay $\dz\to\omega\phi$, which is kinematically allowed, can occur via internal emission of a $W^+$ boson.
The branching fraction (BF) of $\dz\to\omega\phi$ is predicted to be at the level of $10^{-3}$-$10^{-4}$ by various phenomenological models~\cite{kamal, Cheng, kamal, Jiang, Bajc},
and its polarization features are particularly attractive.
No signal for $\dz\to\omega\phi$ has yet been observed, and only an upper limit on the BF, $\mathcal{B} (\Dz \to \omega \phi) < 2.1 \times 10^{-3}$~\cite{albrecht}, is available. 
An observation of $\dz \to \omega \phi$ is desired to shed light on the polarization puzzle, test different theoretical models~\cite{kamal, Bedaque, Hinchliffe},  measure $CP$-violating parameters and strong phases~\cite{Kang, Charles}, and explore the dynamics of $D^0-\bar{D}^0$ mixing~\cite{Falk, Cheng, Jiang}.
It is worth noting that the narrow widths of the $\omega$ and $\phi$ signals allow for a straightforward signal extraction in the process $\Dz\to \omega\phi$, which is unlike other $\Dz\to VV$ decay modes, such as $\Dz\to\rho^0\rho^0$, which require complicated and model dependent analyses of multi-body decays.

In this Letter, we present the first measurement of $\dz \to \omega \phi$ using a $\psipp$ data sample corresponding to an integrated luminosity of 2.93~\ifb~collected by the BESIII detector~\cite{psipp}. The measurement is performed using a single-tag technique, where only one $\dz$ meson in the $\psipp \to \DzDzbar$ decays is reconstructed.
Thus, the BF of $\dz \to \omega \phi$ is calculated by
\begin{equation}
\mathcal{B}= \frac{N_{\rm sig}}{2\cdot N_{\DzDzbar} \cdot \epsilon \cdot \mathcal{B}_{\rm sub}},
\label{BF}
\end{equation}

\noindent where $N_{\rm sig}$ is the signal yield extracted from data, $N_{\DzDzbar}=(10597 \pm 28 \pm 89)\times 10^3$ is the total number of $\psipp\to \DzDzbar$ decays quoted from Ref.~\cite{hajime}, $\epsilon$ is the detection efficiency, and $\mathcal{B}_{\rm sub}$ is the product of BFs for the intermediate-state decays.

In the decay $\dz\to\omega \phi$, polarization amplitudes can be extracted from angular distributions. The differential decay width is given by
\begin{align}
\frac{1}{\Gamma}\frac{d^2 \Gamma}{d\cos\theta_{\omega}d\cos\theta_{K}} =  \frac{9}{4}\{\frac{1}{4}(1-f_L)\sin^2\theta_{\omega}\sin^2\theta_{K}  \nonumber \\
 + f_L\cos^2\theta_{\omega}\cos^2\theta_{K}\},
\label{Eq:helicity}
\end{align}
\noindent where $\theta_{\omega}$ is the angle between $\textbf{p}_{\pi^+}^{\omega} \times \textbf{p}_{\pi^-}^{\omega}$ and $-\textbf{p}_{D^0}^{\omega}$ in the $\omega$ rest frame, 
and $\theta_{K}$ is the angle between $\textbf{p}_{K^-}^{\phi}$ and $-\textbf{p}_{D^0}^{\phi}$ in the $\phi$ rest frame. 
Here, $\textbf{p}_{\pi^+}^{\omega}$, $\textbf{p}_{\pi^-}^{\omega}$, $\textbf{p}_{K^-}^{\phi}$, and $\textbf{p}_{D^0}^{\omega/\phi}$ are the momenta of the $\pi^+, \pi^-, K^-$ and $D^0$, in the rest frame of either the $\omega$ or $\phi$ meson, respectively.
By integrating over $\cos\theta_\omega$ or $\cos\theta_K$ from $-1$ to $+1$, Eq.~\eqref{Eq:helicity} is simplified to be
\begin{equation}
\frac{1}{\Gamma}\frac{d \Gamma}{d\cos\theta} =  \frac{3}{2} \{\frac{1}{2}(1-f_L)\sin^2\theta + f_L\cos^2\theta\},
\label{Eq:fl}
\end{equation}
\noindent where $\theta$ can be either $\theta_{\omega}$ or $\theta_{K}$. 
A detailed illustration of the decay topology, which shows the definitions of the angles can be found in Fig. 1 in Ref.~\cite{supplementary}.

A detailed description of the design and performance of the BESIII detector can be found in Ref.~\cite{besdet}.
A Monte Carlo (MC) simulation tool based on {\sc Geant4}~\cite{geant4} is implemented,
in which the $\epem$ annihilation is simulated with the {\sc KKMC} generator~\cite{2001SJadach} incorporating the effects of beam-energy spread and initial-state-radiation (ISR). An inclusive MC sample, composed of the production of $\DDbar$ pairs, the  non-$\DDbar$ decays of the  $\psi(3770)$, continuum processes of $\epem \to q\bar{q}~(q=u,d,s)$, and the ISR production of the $J/\psi$ and $\psi(3686)$ states,
is used to study the potential background. In the MC sample, the known decay modes are generated with {\sc EvtGen}~\cite{evtgen} using BFs from the Particle Data Group (PDG)~\cite{pdg}, and the remaining unknown decays are generated with {\sc Lundcharm}~\cite{lundcharm}.
The signal sample of $\Dz \to \phi \omega$ decays is modeled by a pseudoscalar meson decaying into two vector mesons  with  transverse polarization using {\sc EvtGen}~\cite{evtgen}.

The $\phi$ and $\omega$ candidates are reconstructed from their dominant decays $\phi\to \Kp\Km$ and $\omega\to \pip\pim\piz$, respectively, where the $\piz$ is identified by a photon pair. The charged tracks must be within the main drift chamber (MDC) acceptance region by requiring the polar angle $|\cos\theta| < 0.93$,
and must originate from the interaction point (IP) with a distance of closest approach within $\pm 1$ cm in the plane perpendicular to the beam and $\pm 10$ cm along the beam direction. Particle identification (PID) is performed by requiring $\mathcal{L}_{\pi}>\mathcal{L}_K$ and $\mathcal{L}_K > \mathcal{L}_{\pi}$ for the $\pi^{\pm}$ and $K^{\pm}$ candidates, respectively, where $\mathcal{L}_{\pi}$ and $\mathcal{L}_{K}$ are the likelihoods for the pion and kaon hypotheses calculated by combining the time-of-flight (TOF) information from the TOF detector and the $dE/dx$ information from the MDC.

Photon candidates are selected from neutral showers deposited in the electromagnetic calorimeter (EMC) with energies larger than 25~MeV in the barrel region ($|\cos\theta| < 0.80$) and  50~MeV in the end-cap regions ($0.86 < |\cos\theta| < 0.92$). The EMC timing is required to be within 700~ns relative to the event start time to suppress electronic noise and deposited energy unrelated to the collision events.
Furthermore, a photon candidate is required to be at least $10^{\degree}$ away from any charged tracks to  avoid any overlap between them. A $\piz$ candidate is formed by a photon pair with invariant mass within  $(0.115, 0.150)$~GeV/$c^2$.
To improve the resolution, a kinematic fit is imposed on the selected photon pair by constraining their invariant mass at the nominal $\piz$ mass~\cite{pdg}, and the resultant kinematic variables are used in the subsequent analysis.

To identify the $\Dz$ signal, the energy difference $\De = E_D-\Ebeam$ and the beam-constrained mass $\mbc= \sqrt{\Ebeam^2/c^4-p_D^2/c^2}$ are calculated, where $\Ebeam$ is the beam energy, and $E_D$ ($p_D$) is the reconstructed energy (momentum) of the $\Dz$ candidate in the $\epem$ center-of-mass system. The $\Dz$ signal peaks around zero in the $\De$ distribution and around the nominal $\Dz$  mass ($m_D$) in the $\mbc$ distribution. The  $\Dz\to\omega\phi$ signal is reconstructed from all possible $\pppm\piz\Kp\Km$ combinations. 
If there is more than one combination, the one with a minimum value of $|\De|$ is selected. A $\Dz$ candidate is required to satisfy $\mbc >1.84~\gevcc$ and $-0.03<\De<0.02~\gev$. The $\De$ requirement corresponds to an interval of 4 standard deviations from the peak position. The asymmetric boundaries stem from the photon energy detection in the EMC. A prominent peak corresponding to the $K_S^0$  in the $M_{\pi^+\pi^-}$ distribution, arising from the background process $D^0 \to K_S^0+{\rm anything}$, is rejected by removing the mass range $(0.490, 0.503)$~\gevcc.  

 Figure~\ref{Mbcdist} shows the $\mbc$ distribution for the surviving events of data and the background predictions from various MC samples with the $\Kp\Km$ invariant mass $M_{\Kp\Km}<1.05~\gevcc$ and the $\pppm\piz$ invariant mass $M_{\pppm\piz}>0.65~\gevcc$, where the clear peak around $m_D$ in data refers to the signal of $\Dz\to\pppm\piz\Kp\Km$.

 \begin{figure}[htbp]

  \centering
 \includegraphics[width=0.4\textwidth]{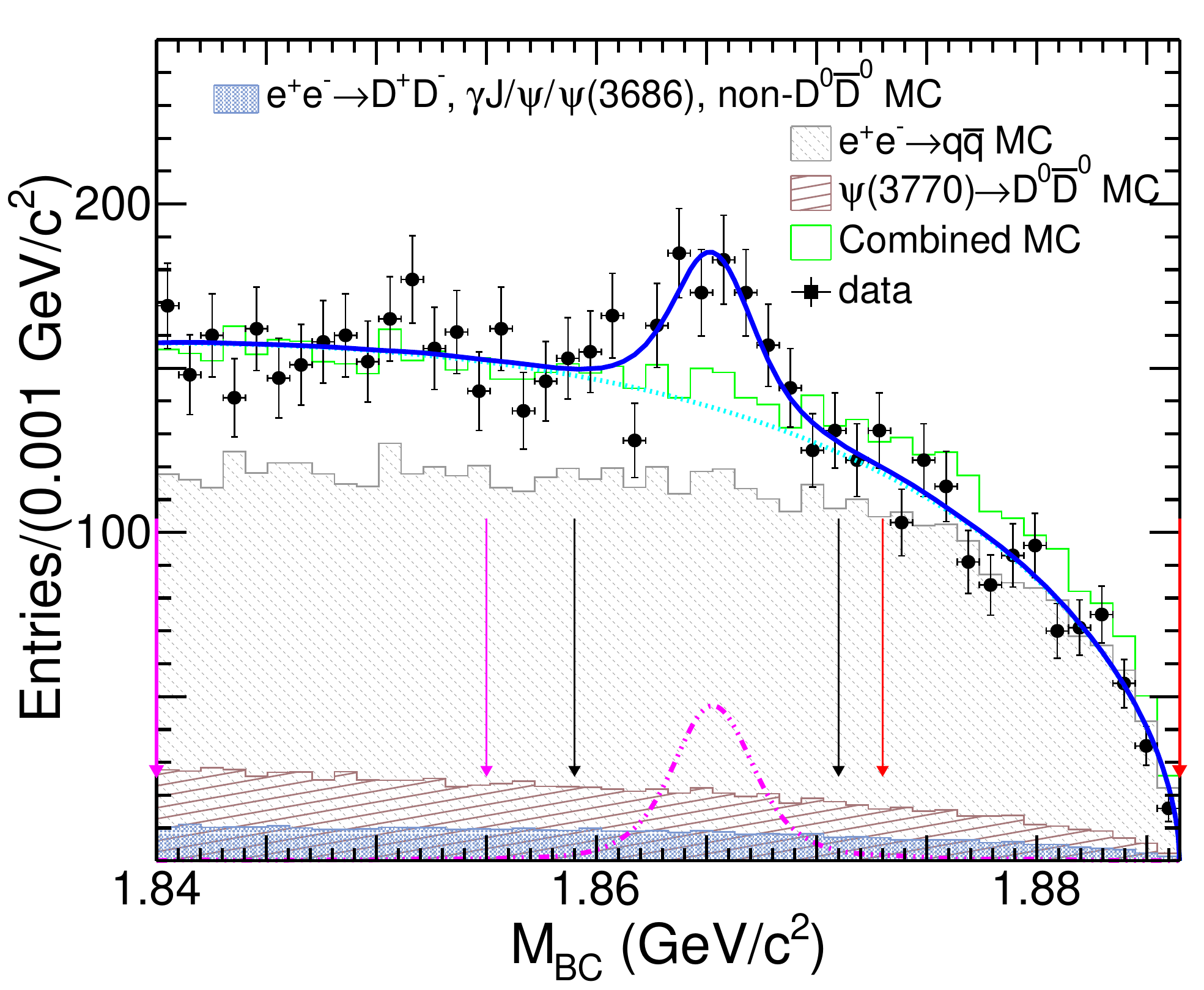}

 \caption{Fit to the $\mbc$ distribution of the candidate events for $D^0 \to \pi^+\pi^-\pi^0 K^+ K^-$. Black dots with error bars are data, dashed cyan curve for combinatorial background, long dashed-dotted pink curve for the $\dz$ signal, the solid blue curve for the total fit, and shadow histograms for the non-$D^0$ background predictions from various MC samples. The two black and two pink (red) arrows  represent the $\mbc$ signal and low (high)-sideband regions, respectively.}
\label{Mbcdist} 
\end{figure}

The $\Dz\to\omega\phi$ signal is evident in
Fig.~\ref{2D_projplot1}, where
the  distribution of $M_{\pppm\piz}$ versus $M_{\Kp\Km}$ as well as their corresponding projection plots are shown  
for events in the $\mbc$ signal region ($1.859, 1.871)~\gevcc$ and sideband region ($1.840, 1.855) \cup (1.873,1.890)~\gevcc$. A cluster of events around the intersection of the $\omega$ and $\phi$ nominal masses in the $\mbc$ signal region indicates the signal $\dz\to\omega\phi$.  There is no corresponding cluster of events in the sideband plot. Clear $\phi$ signal events are observed in the $\mbc$ sideband region, indicating the contribution of the $\phi$ meson from non-$\dz$ decays.
Prominent $\omega$ signal events are present in the $\mbc$ signal region but absent in the corresponding sideband region, indicating the contribution of the $\omega$ meson from $D^0$ decays.

\begin{figure}[htbp]
  \centering
 \includegraphics[width=0.50\textwidth]{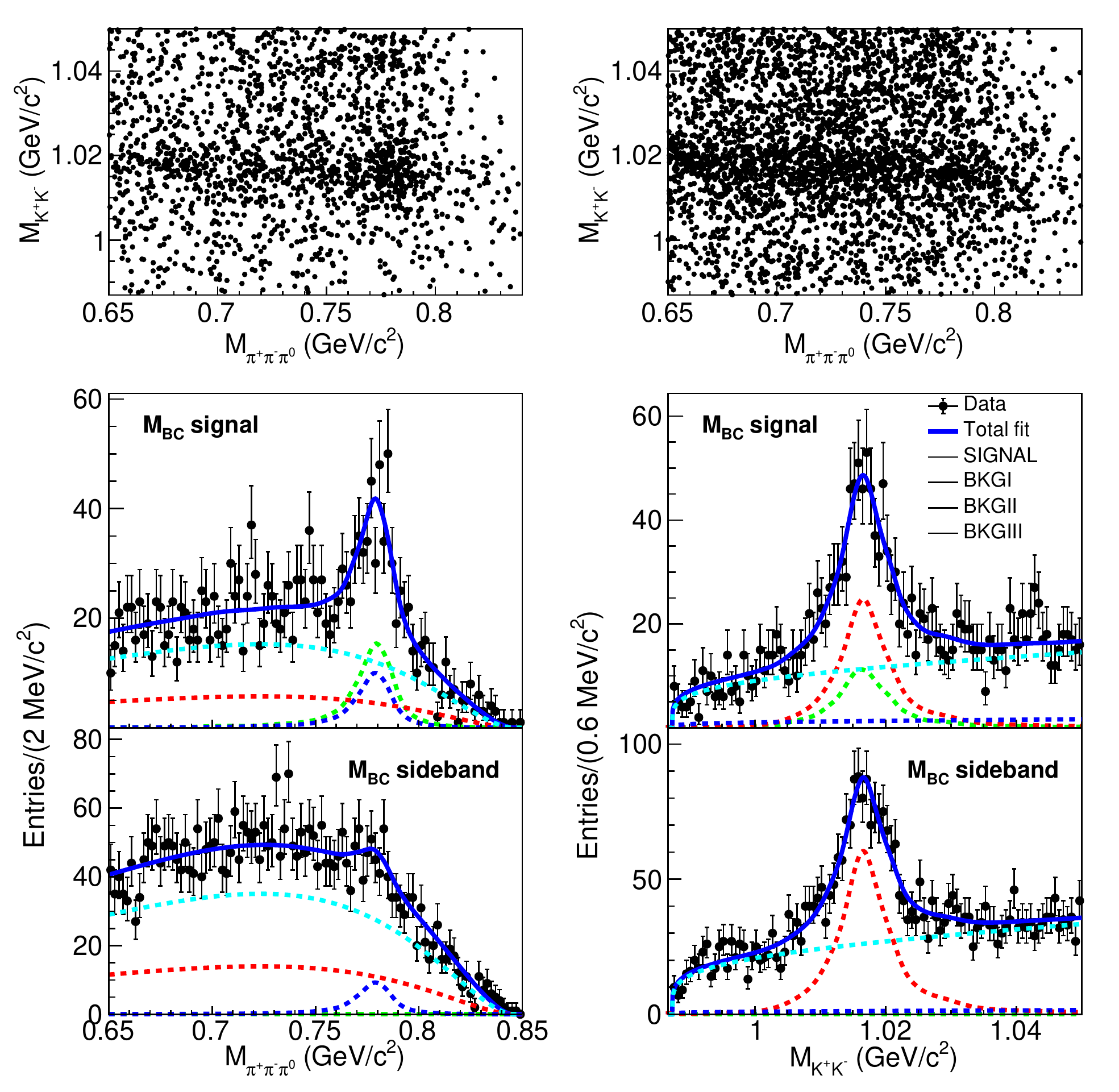}

 \caption{(Top) the distributions of $M_{\Kp\Km}$  versus $M_{\pppm\piz}$   in the $\mbc$ signal (left) and sideband (right) regions, and (middle and bottom) the corresponding 1-D projection plots of $M_{\pi^+\pi^-\pi^0}$ (left)  and $M_{K^+K^-}$ (right). 
 In the projection plots, the black dots with error bars are data, the solid blue, dashed red,  dotted green, dashed-dotted blue, and long dashed-dotted cyan curves represent total fit results, SIGNAL, BKGI, BKGII, and BKGIII, respectively.}
  \label{2D_projplot1}
\end{figure}

To extract the signal yield, a two-dimensional (2D) unbinned maximum likelihood fit is performed on the $M_{\pppm\piz}$ versus $M_{\Kp\Km}$ distributions.  This fit is performed simultaneously in both the $\mbc$ signal and sideband regions, where the sideband events are used to constrain the background from  non-$\dz$ decays.
The fit includes a signal component, SIGNAL, which has both $\omega$ and $\phi$ intermediate states, and three backgrounds, BKGI, BKGII, and BKGIII. The BKGI (BKGII) contains only the $\omega$ ($\phi$) intermediate state, and BKGIII includes neither the $\omega$ nor $\phi$ intermediate states. It is worth noting that the above four components may exist in both $\dz$ and non-$\dz$ decays. The yield of the signal $\dz\to\omega\phi$ is extracted from the $\mbc$ signal region by subtracting the contribution from non-$\dz$ decays estimated from the $\mbc$ sideband region.

The SIGNAL is described by a distribution obtained from a 2D kernel estimation~\cite{kernal} of the unbinned signal MC samples.
BKGI is parameterized with the product of a distribution obtained from a 1D kernel estimation~\cite{kernal} of the $\omega$ signal MC for the $M_{\pppm\piz}$ distribution and a reversed ARGUS function~\cite{argus}  defined by the formula of Eq.(4) in Ref.~\cite{invisible} for the $M_{\Kp\Km}$ distribution. Vice versa, BKGII is described with the product of an ARGUS function for the $M_{\pppm\piz}$ distribution and a distribution obtained from a 1D kernel estimation of the $\phi$ signal MC for the $M_{\Kp\Km}$ distribution. BKGIII is the product of an ARGUS function for the $M_{\pppm\piz}$ distribution and a reversed ARGUS function for the $M_{\Kp\Km}$ distribution. 
To compensate for the resolution difference between data and simulation, the shapes derived from simulation are convolved with (1D or 2D) Gaussian functions, which share the same parameters between different fit components and these parameters are floated during the fit. The endpoints of the ARGUS functions are fixed to the corresponding threshold values of $(m_{D}-m_{\phi})$ and $2m_{K^{\pm}}$, respectively, where $m_{\phi}$ ($m_{K^{\pm}}$) is the nominal mass of the $\phi$ ($K^{\pm}$) meson~\cite{pdg}.

Detailed MC studies show that the non-peaking background shapes in the $M_{\Kp\Km}$ distributions are identical in both the $\mbc$ signal and sideband regions, but slightly different for $M_{\pppm\piz}$ distributions due to the threshold effect of kinematics. Thus, the reversed ARGUS parameterizations of the  $M_{\Kp\Km}$ distributions share the same parameters in both $\mbc$ signal and sideband regions, but no constraint is implemented for the  ARGUS functions for the $M_{\pppm\piz}$ distributions in different $\mbc$ regions. 
We float SIGNAL, BKGI, BKGII, and BKGIII components in both $\mbc$ signal and sideband regions during the fit. The final signal yield is also constrained to be $N_{\rm SG}= N_{\rm sig}+f\cdot N_{\rm SB}$, where $N_{\rm SG}$ and $N_{\rm SB}$ are the numbers of the SIGNAL component in 
the $\mbc$ signal and sideband regions, respectively, as shown in Fig.~\ref{2D_projplot1}.
 The factor $f$ is the ratio of the corresponding yields from the non-$\dz$ decay in the $\mbc$ signal and sideband regions, and its value is determined to be  $(44.3 \pm 0.9)\%$ by fitting the $\mbc$ distribution, as shown in Fig.~\ref{Mbcdist}. In this fit, the $\dz$ signal is described by the simulated signal shape convolved with a Gaussian function while the non-$\dz$ background by an ARGUS function~\cite{argus}, where we fix signal shape and float rest of the other parameters during the fit. The 2D simultaneous fit yields $N_{\rm sig}=195.9\pm29.1$, which includes the uncertainties from $N_{\rm SB}$ and $N_{\rm SG}$. The detection efficiency is calculated to be $(3.32\pm0.04)\%$ by the same 2D simultaneous fit approach with an inclusive MC sample, which is a mixture of the signal MC sample generated by considering the polarization of $\dz\to\omega\phi$ as discussed below, and various backgrounds. The BF of $\dz\to\omega\phi$ is determined to be 
$(6.48 \pm 0.96 \pm 0.40)\times 10^{-4}$ according to Eq.~\eqref{BF}, where the first and second uncertainties are statistical and systematic, respectively. The corresponding significance is 6.3~$\sigma$ calculated by $\sqrt{-2\ln (\mathcal{L}_0/\mathcal{L}_{\rm max})}$ including both statistical and systematic unncertainties,
where $\mathcal{L}_{\rm max}$ and $\mathcal{L}_0$ are the likelihood values for the nominal fit and the alternative fit with zero signal assumption, respectively. Different contributions to the systematic uncertainty will be described later.

To study the polarization in the $\dz\to\omega\phi$ decay, the efficiency-corrected signal yields are evaluated in five equal bins of $|\cos\theta_{\omega}|$ and $|\cos\theta_K|$ as shown in Fig.~\ref{efficiencycorNsig}. Here, we extract the signal yield in each bin using a procedure similar to the 2D simultaneous fit approach discussed above.  The corresponding detection efficiency is obtained using a simulated signal sample generated uniformly over phase space (PHSP). A joint $\chi^2$ fit on the $|\cos\theta_{\omega}|$ and $|\cos\theta_K|$ distributions of data is performed  with Eq.~\eqref{Eq:fl}, where $f_L$ is floated between $[-1,1]$. The fit yields $f_L=0.00 \pm 0.10 \pm 0.08$, which corresponds to $f_L< 0.24$ at $95\%$ confidence level computed by integrating the likelihood versus $f_L$ curve from zero to $95\%$ of the total curve after including the systematic uncertainty as described below. This result indicates that the vector mesons are transversely polarized in the $\dz\to\omega\phi$ decay. 

\begin{figure}[htbp]
  \centering

 \includegraphics[width=0.5\textwidth]{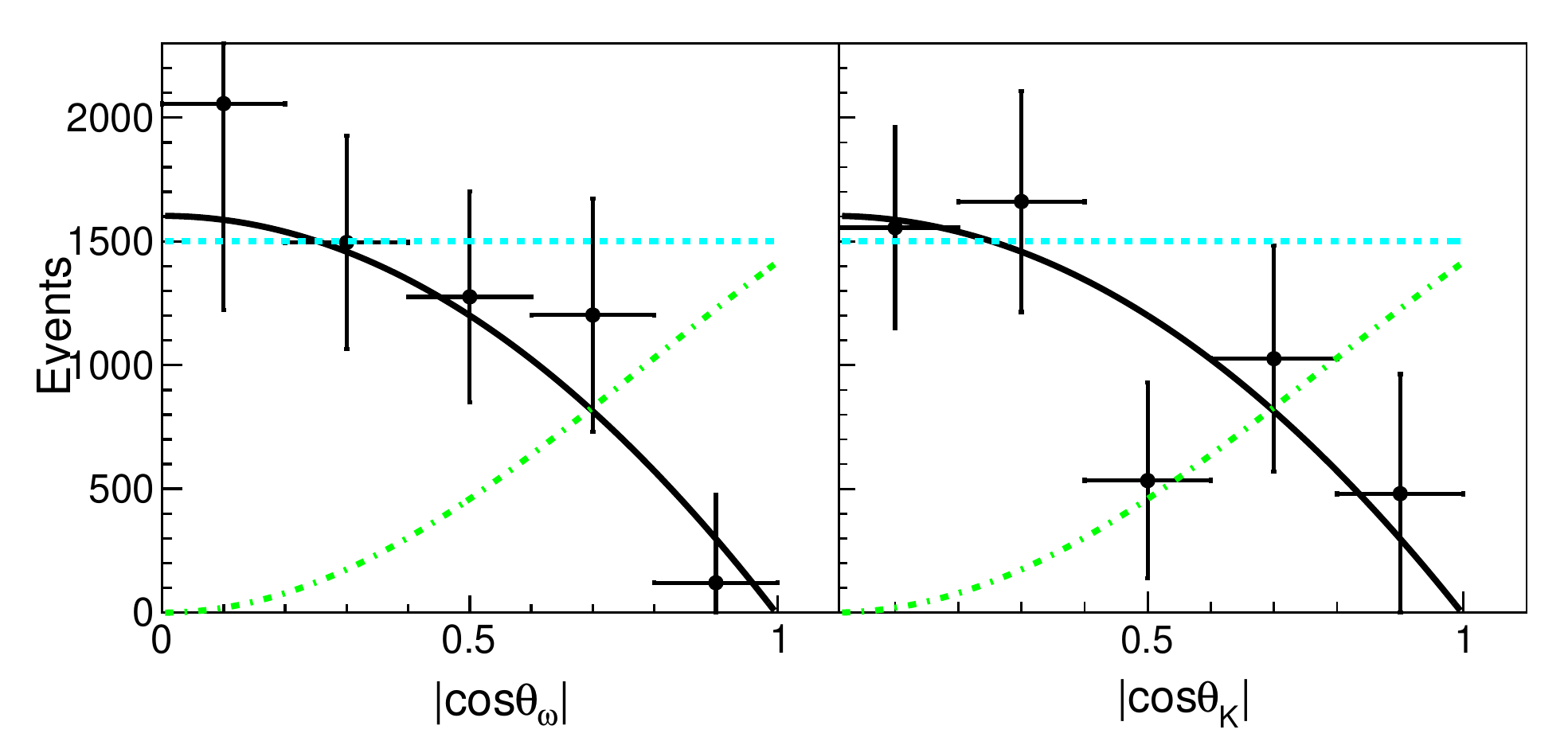}

  \caption{The distributions of the background-subtracted signal yields corrected by the efficiency versus $|\cos\theta_{\omega}|$ (left) and $|\cos\theta_K|$ (right). The black dots with error bars are data with both statistical and systematic uncertainties,  
  and the solid black curves are the fit results. 
  The distributions with the longitudinal polarization  and PHSP assumptions are  shown as the 
  dotted dashed green and dashed cyan curves, respectively. 
   }
  \label{efficiencycorNsig}
\end{figure}

According to Eq.~\eqref{BF}, the systematic uncertainties for the BF measurement include those from the reconstruction efficiency,  MC modeling, signal yield, number of $\DzDzbar$ events, and the BFs of the intermediate-state decays.
The systematic uncertainties associated with the reconstruction efficiency include tracking and PID of the charged tracks, $\piz$ reconstruction, $\Delta E$ requirement, and $\ks$ veto.
The systematic uncertainties from the MC modeling include those from the MC statistics, $\omega \to \pi^+\pi^-\pi^0$ modeling, quantum correlation (QC) effect, and the longitudinal polarization fraction $f_L$, which varies from $f_L=0$ to $1\sigma$ upper bound uncertainty of $0.13$. 
The systematic uncertainty due to the 2D simultaneous fit includes those from signal and background probability density functions (PDFs), the ratio of background between the $\mbc$ signal and sideband regions ($f$),  and the fit bias.
All of the above systematic uncertainties are estimated with different approaches. Refer to the supplementary material~\cite{supplementary} for details.
The uncertainties of $N_{D^0\bar{D}^0}$ and the BFs of the intermediate-state decays are from Ref.~\cite{hajime} and PDG~\cite{pdg}, respectively.
The total systematic uncertainty is $6.2\%$ calculated by summing all individual uncertainties quadratically and assuming them to be independent.

The systematic uncertainty for the $f_L$ measurement includes those from MC modeling, $M_{\rm BC}$ signal region, background fraction $f$, different bin size of $\cos\theta_{\omega, K}$, and signal and background PDFs. We replace the PHSP signal sample  with a MC sample generated under the hypothesis of transverse polarization to evaluate the efficiency in each bin of the $\cos\theta_{\omega}$ and $\cos\theta_K$ distributions. We also extract the signal yields with the alternative $M_{\rm BC}$ signal region enlarged by $2$ MeV/$c^2$, background fraction $f$, different bin size of $\cos\theta_{\omega}$ and $\cos\theta_K$, and signal and background PDFs as done for the BF measurement. A joint $\chi^2$ fit is performed to each set of the efficiency corrected signal yields versus $\cos\theta_{\omega}$ and $\cos\theta_{K}$ data, and the resultant change in $f_L$ is considered as a systematic uncertainty. The total systematic uncertainty is 0.08 calculated as the quadratic sum of the individual ones.

In summary, the decay $\dz\to\omega\phi$ is observed for the first time with a significance of $6.3\sigma$ by analyzing the $\psipp$ data taken by the BESIII experiment, corresponding to an integrated luminosity of 2.93~\ifb.
The measured BF is  $(6.48 \pm 0.96 \pm 0.40)\times 10^{-4}$, which is consistent with the factorization model predictions~\cite{kamal, Cheng}, but inconsistent with predictions based on SU(3) symmetry with nonet symmetry~\cite{kamal}, the factorization-assisted topological-amplitude method~\cite{Jiang}, and the heavy quark effective Lagrangian and chiral perturbation theory~\cite{Bajc}.
Our angular distribution studies reveal that the $\omega$ and $\phi$ in the decay $\dz\to\omega\phi$ are transversely polarized, which is the same as that observed in the decay $\dz\to\bar{K}^{*0}\rho^0$, but contradicts predictions from the naive factorization~\cite{Cheng} and Lorentz invariant-based symmetry~\cite{Gudrun} models. 
The results challenge our understainding of the underlying dynamics in charmed meson decays, and also may help in searches for new physics. 

The BESIII collaboration thanks the staff of BEPCII, the IHEP computing center and the supercomputing center of USTC for their strong support. 
Authors are grateful to Dr. Fusheng Yu, Dr. Yuelong Shen and Dr. Xiangwei Kang for their enlightening discussions.
This work is supported in part by National Key Research and Development Program of China under Contracts Nos. 2020YFA0406400, 2020YFA0406300; National Natural Science Foundation of China (NSFC) under Contracts Nos. 11625523, 11635010, 11735014, 11822506, 11835012, 11935015, 11935016, 11935018, 11961141012, 12022510, 12035009, 12035013, 12061131003, 11605196, 11605198, 11705192, 11950410506; $64^{th}$ batch of Postdoctoral Science Fund Foundation under contract No. 2018M642516; the Chinese Academy of Sciences (CAS) Large-Scale Scientific Facility Program; Joint Large-Scale Scientific Facility Funds of the NSFC and CAS under Contracts Nos. U1732263, U1832207, U1832103, U2032111; CAS Key Research Program of Frontier Sciences under Contract No. QYZDJ-SSW-SLH040; 100 Talents Program of CAS; INPAC and Shanghai Key Laboratory for Particle Physics and Cosmology; ERC under Contract No. 758462; European Union Horizon 2020 research and innovation programme under Contract No. Marie Sklodowska-Curie grant agreement No 894790; German Research Foundation DFG under Contracts Nos. 443159800, Collaborative Research Center CRC 1044, FOR 2359, FOR 2359, GRK 214; Istituto Nazionale di Fisica Nucleare, Italy; Ministry of Development of Turkey under Contract No. DPT2006K-120470; National Science and Technology fund; Olle Engkvist Foundation under Contract No. 200-0605; STFC (United Kingdom); The Knut and Alice Wallenberg Foundation (Sweden) under Contract No. 2016.0157; The Royal Society, UK under Contracts Nos. DH140054, DH160214; The Swedish Research Council; U. S. Department of Energy under Contracts Nos. DE-FG02-05ER41374, DE-SC-0012069.



\end{document}




\begin{center}
\textbf{\boldmath  \LARGE{First measurement of polarization in $\dz\to\omega\phi$ decay:\\
                 Supplemental material}}
\end{center}

\Large

\section{\boldmath \large The decay topology of $\dz\to\omega\phi$ and the defination of the decay angles}

\begin{figure}[htbp]
 \centering
 \includegraphics[width=0.45\textwidth]{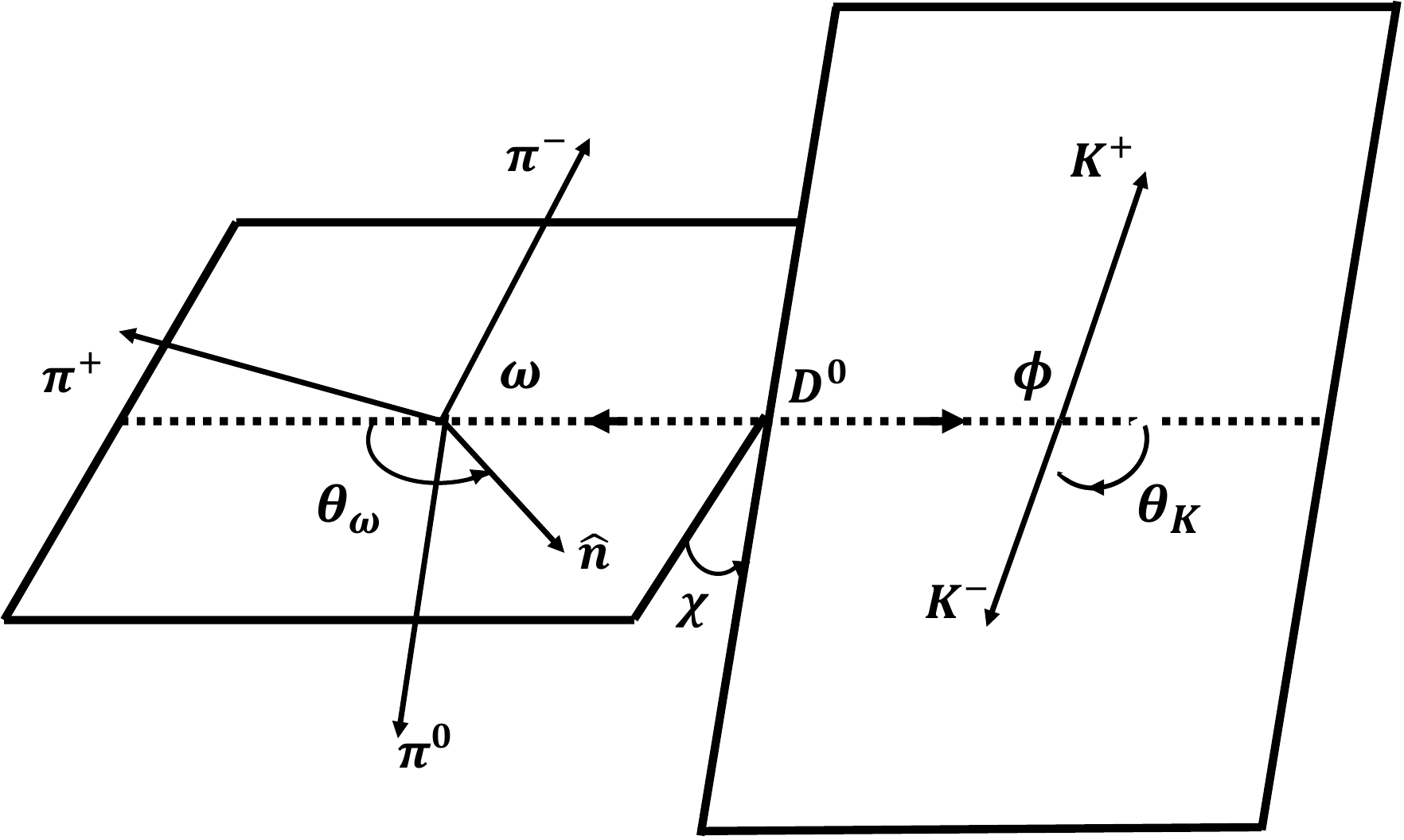}%

 \caption{
 The decay topology of $\dz\to\omega\phi$ and the definitions of the decay angles.}

  \label{helicity}
\end{figure}

Figure~\ref{helicity} illustrates the decay topology of $\dz\to\omega\phi$ as well as the definitions of $\theta_{\omega}$ and $\theta_{\phi}$ using in the polarization analysis.

\section{\boldmath \large The detail uncertainties associated with the reconstruction effiency}

The uncertainties associated with the reconstruction efficiency include tracking and PID of the charged tracks, $\piz$ reconstruction, $\Delta E$ requirement, and $\ks$ veto.

The uncertainty associated with the tracking efficiency is studied using a control sample of $\psipp \to \DDbar$ with hadronic $D$ decays via a partial reconstruction method~\cite{ST1, ST2}, where a small deviation between data and simulation is present for kaon tracks with momenta less than 0.35~GeV/$c$.
The kaons from $\phi$ decay in the signal are of low momenta. Consequently, a correction factor of 1.06 for $K^+K^-$ is applied in the detection efficiency, and an uncertainty of 0.5\% is assigned for each kaon or pion.
The correction factor is the ratio of the efficiencies of data and simulation weighted according to the kaon momentum distribution. We also utilize this control sample to compute the uncertainties associated with PID ($0.5\%$) and $\piz$ reconstruction efficiency ($2.0\%$)~\cite{tracking}.

The uncertainty originating from the $\De$ requirement is studied using a control sample of $\dz\to 2(\pppm)\piz$ decays, which has a similar final state as the signal except with a pion pair instead of a kaon pair.
The control sample is selected by a relatively loose $\De$ requirement, $i.e.$, $\De<0.1~\gev$,  
and the corresponding signal yield is extracted by fitting the $\mbc$ distribution.
The nominal $\De$ requirement is then implemented on the control sample, and the resultant ratio of signal yields is taken as the efficiency.
The approach is implemented for both data and inclusive MC samples, and the resultant difference in the data and MC efficiencies, 1.4\%, is taken as the uncertainty.

The uncertainty from the $\ks$ veto is studied by varying the $\ks$ mass window requirement within $\pm 1\sigma$,
and the larger difference in the BF, 0.8\%, is taken as the uncertainty.

The total  uncertainties associated with the reconstruction efficiency is $3.8\%$, which is the quadratic sum of above individual ones.

\section{\boldmath \large The detail uncertainties associated with the MC modeling}

The uncertainties from the MC modeling includes those from the MC statistics ($0.8\%$), $\omega \to \pi^+\pi^-\pi^0$ modeling, quantum correlation (QC)~\cite{QC} effect, and the longitudinal polarization fraction $f_L$. The uncertainty due to the $\omega \to \pi^+\pi^-\pi^0$ modeling is assigned to be $0.5\%$ on the basis of two MC samples generated with two different models~\cite{Wasa, BESIIIDP}.
From the analysis, the decay $\dz\to\omega\phi$ appears to be transversely polarized, thus it is a mixture of $CP$-even and $CP$-odd components.
The uncertainties associated with the polarization is studied by an alternative signal MC sample generated with $1\sigma$ upper bound uncertainty, $f_L=0.13$, and the resultant change in the efficiency, 3.2\%, is taken as the uncertainty.

The total uncertainties associated with the MC modeling is $3.3\%$, which is the quadratic sum of above individual ones.

\section{\boldmath \large The detail uncertainties associated with 2D simultaneous fits}
The systematic uncertainty due to the 2D simultaneous fit includes those from signal and background probability density functions (PDFs), the ratio of background between the $\mbc$ signal and sideband regions ($f$),  and the fit bias.
The uncertainty arising from the signal PDF, 1.2\%, is evaluated with an alternative fit, in which the signal PDFs are described using a different non-parameterized modeling of the simulated shape, convolved with a Gaussian function.
The uncertainty of the background PDF, 0.4\%, is determined by replacing the ARGUS function~\cite{argus} with a modified one as used in Ref.~\cite{invisible}.
The uncertainty from $f$ is $0.1\%$, evaluated by varying its value within 1~$\sigma$ when calculating the signal yield.
The uncertainty due to the choice of the $\mbc$ signal region is evaluated to be $2.7\%$ by enlarging its region by 2~$\mevcc$, which is the resolution of the $M_{\rm BC}$ distribution. The fit bias, 1.0\%, is estimated with a large number of pseudo-experiments. Each pseudo-experiment sample is a composition of the signal generated according to the signal PDF and background expectations from the inclusive MC sample. The resultant pull distribution for the BF is consistent with a normal distribution, and we consider the average fit bias as the uncertainty.

The total uncertainty associated with the 2D simultaneous fits is $3.2\%$, which is the quadratic sum of above individual ones.